\documentclass[aps,prl,a4paper,superscriptaddress,twocolumn,showpacs,amsmath,amssymb,longbibliography]{revtex4-2}

\usepackage{graphicx,epsfig}
\usepackage[usenames]{color}

\usepackage[english]{babel}

\usepackage{dsfont}

\usepackage{csquotes}
\MakeOuterQuote{"}

\allowdisplaybreaks

\begin{document}

\newcommand{\E}{\mathcal{E}}
\newcommand{\G}{\mathcal{G}}
\newcommand{\Lag}{\mathcal{L}}
\newcommand{\M}{\mathcal{M}}
\newcommand{\N}{\mathcal{N}}
\newcommand{\U}{\mathcal{U}}
\newcommand{\R}{\mathcal{R}}
\newcommand{\F}{\mathcal{F}}
\newcommand{\V}{\mathcal{V}}
\newcommand{\C}{\mathcal{C}}
\newcommand{\I}{\mathcal{I}}
\newcommand{\s}{\sigma}
\newcommand{\up}{\uparrow}
\newcommand{\dw}{\downarrow}
\newcommand{\h}{\hat{\mathcal{H}}}
\newcommand{\himp}{\hat{h}}
\newcommand{\g}{\mathcal{G}^{-1}_0}
\newcommand{\D}{\mathcal{D}}
\newcommand{\A}{\mathcal{A}}
\newcommand{\projs}{\hat{\mathcal{S}}_d}
\newcommand{\proj}{\hat{\mathcal{P}}_d}
\newcommand{\K}{\textbf{k}}
\newcommand{\Q}{\textbf{q}}
\newcommand{\T}{\tau_{\ast}}
\newcommand{\io}{i\omega_n}
\newcommand{\eps}{\varepsilon}
\newcommand{\+}{\dag}
\newcommand{\su}{\uparrow}
\newcommand{\giu}{\downarrow}
\newcommand{\0}[1]{\textbf{#1}}
\newcommand{\ca}{c^{\phantom{\dagger}}}
\newcommand{\cc}{c^\dagger}

\newcommand{\pa}{{p}^{\phantom{\dagger}}}
\newcommand{\pc}{{p}^\dagger}

\newcommand{\fa}{f^{\phantom{\dagger}}}
\newcommand{\fc}{f^\dagger}
\newcommand{\aaa}{a^{\phantom{\dagger}}}
\newcommand{\aac}{a^\dagger}
\newcommand{\bba}{b^{\phantom{\dagger}}}
\newcommand{\bbc}{b^\dagger}
\newcommand{\da}{{d}^{\phantom{\dagger}}}
\newcommand{\dc}{{d}^\dagger}
\newcommand{\ha}{h^{\phantom{\dagger}}}
\newcommand{\hc}{h^\dagger}
\newcommand{\be}{\begin{equation}}
\newcommand{\ee}{\end{equation}}
\newcommand{\bea}{\begin{eqnarray}}
\newcommand{\eea}{\end{eqnarray}}
\newcommand{\ba}{\begin{eqnarray*}}
\newcommand{\ea}{\end{eqnarray*}}
\newcommand{\dagga}{{\phantom{\dagger}}}
\newcommand{\bR}{\mathbf{R}}
\newcommand{\bQ}{\mathbf{Q}}
\newcommand{\bq}{\mathbf{q}}
\newcommand{\bqp}{\mathbf{q'}}
\newcommand{\bk}{\mathbf{k}}
\newcommand{\bh}{\mathbf{h}}
\newcommand{\bkp}{\mathbf{k'}}
\newcommand{\bp}{\mathbf{p}}
\newcommand{\bL}{\mathbf{L}}
\newcommand{\bRp}{\mathbf{R'}}
\newcommand{\bx}{\mathbf{x}}
\newcommand{\by}{\mathbf{y}}
\newcommand{\bz}{\mathbf{z}}
\newcommand{\br}{\mathbf{r}}
\newcommand{\Ima}{{\Im m}}
\newcommand{\Rea}{{\Re e}}
\newcommand{\Pj}[2]{|#1\rangle\langle #2|}
\newcommand{\ket}[1]{\vert#1\rangle}
\newcommand{\bra}[1]{\langle#1\vert}
\newcommand{\setof}[1]{\left\{#1\right\}}
\newcommand{\fract}[2]{\frac{\displaystyle #1}{\displaystyle #2}}
\newcommand{\Av}[2]{\langle #1|\,#2\,|#1\rangle}
\newcommand{\av}[1]{\langle #1 \rangle}
\newcommand{\Mel}[3]{\langle #1|#2\,|#3\rangle}
\newcommand{\Avs}[1]{\langle \,#1\,\rangle_0}
\newcommand{\eqn}[1]{(\ref{#1})}
\newcommand{\Tr}{\mathrm{Tr}}

\newcommand{\Vb}{\bar{\mathcal{V}}}
\newcommand{\Vd}{\Delta\mathcal{V}}
\def\P{P_{02}}
\newcommand{\Pb}{\bar{P}_{02}}
\newcommand{\Pd}{\Delta P_{02}}
\def\t{\theta_{02}}
\newcommand{\tb}{\bar{\theta}_{02}}
\newcommand{\td}{\Delta \theta_{02}}
\newcommand{\Rb}{\bar{R}}
\newcommand{\Rd}{\Delta R}
\newcommand{\ocrev}[1]{{\color{cyan}{#1}}}
\newcommand{\occom}[1]{{\color{red}{#1}}}

\title{Quantum-embedding description of the Anderson lattice model with \\ the ghost Gutzwiller Approximation}

\author{Marius S. Frank}
\affiliation{Department of Physics and Astronomy, Aarhus University, 8000, Aarhus C, Denmark}
\author{Tsung-Han Lee}
\affiliation{Physics and Astronomy Department, Rutgers University, Piscataway, New Jersey 08854, USA}
\author{Gargee Bhattacharyya}
\affiliation{Department of Physics and Astronomy, Aarhus University, 8000, Aarhus C, Denmark}
\author{Pak Ki Henry Tsang}
\affiliation{Department of Physics and National High Magnetic Field Laboratory, Florida State University, Tallahassee, Florida 32306, USA}
\author{Victor L. Quito}
\affiliation{Department of Physics and Astronomy, Iowa State University, Ames, Iowa 50011, USA}
\affiliation{Department of Physics and National High Magnetic Field Laboratory, Florida State University, Tallahassee, Florida 32306, USA}
\author{Vladimir Dobrosavljevi\'c}
\affiliation{Department of Physics and National High Magnetic Field Laboratory, Florida State University, Tallahassee, Florida 32306, USA}
\author{Ove Christiansen}
\affiliation{Department of Chemistry, Aarhus University, 8000, Aarhus C, Denmark}
\author{Nicola Lanat\`a}
\altaffiliation{Corresponding author: lanata@phys.au.dk}
\affiliation{Department of Physics and Astronomy, Aarhus University, 8000, Aarhus C, Denmark}
\affiliation{Nordita, KTH Royal Institute of Technology and Stockholm University, Hannes Alfv\'ens v\"ag 12, SE-106 91 Stockholm, Sweden}

\date{\today}

\begin{abstract}

We present benchmark calculations of the Anderson lattice model based on the recently-developed "ghost Gutzwiller approximation".
%
%
Our analysis shows that, in some parameters regimes, the predictions of the standard Gutzwiller approximation can be incorrect by orders of magnitude for this model.
We show that this is caused by the inability of this method to describe simultaneously the Mott physics and the hybridization between correlated and itinerant degrees of freedom ---whose interplay often governs the metal-insulator transition in real materials.
Finally, we show that the ghost Gutzwiller approximation solves this problem, providing us with results in remarkable agreement with dynamical mean field theory throughout the entire phase diagram, while being much less computationally demanding.
We provide an analytical explanation of these findings and discuss
their implications within the context of ab-initio computation of strongly-correlated matter.

\end{abstract}

\maketitle

Understanding and simulating quantitatively the electronic behavior of strongly correlated matter is one of the most fundamental problems in condensed-matter science.
The substantial progress achieved today in this direction largely owes to quantum embedding methods~\cite{Kotliar-Science,quantum-embedding-review}.
In particular, the development of dynamical mean field theory (DMFT)~\cite{DMFT,dmft_book,CDMFT-Jarrell,CDMFT-Kotliar,CDMFT-Lichtenstein,CDMFT-Potthoff,CDMFT-Senechal,Aichhorn-DMFT-LAPW,LDA+U+DMFT,Held-review-DMFT,xidai_impl_LDA+DMFT} constituted a great leap in our understanding of strong-correlation phenomena, which advanced dramatically our ability of describing the properties of real materials.
In the past decade, the perspective of expanding the predictive power of simulations within the blooming field of theory-assisted materials-by-design~\cite{MaterialsGenome-1,MaterialsGenome-2} contributed to stimulate the development of alternative computational frameworks, capable of taking into account strong correlations at a lower computational cost.
Within this context, particularly promising approaches are the Gutzwiller approximation (GA)~\cite{Gutzwiller3,Fang,Ho,Bunemann-pnictides,Our-PRX,Gmethod} ---or, equivalently~\cite{equivalence_GA-SB,lanata-barone-fabrizio}, the rotationally-invariant slave-boson mean-field theory~\cite{Fresard1992,Georges,Lanata2016} --- and density matrix embedding theory~\cite{DMET,dmet-ghostcopy1}.
These frameworks have similar algorithmic structures. 
In fact, as in density matrix embedding theory, the GA equations can be cast in terms of ground-state calculations of auxiliary impurity models called embedding Hamiltonians (EH), where the bath has the same number of degrees of freedom as the impurity~\cite{Our-PRX}.
Furthermore, density matrix embedding theory can be formally derived from the GA equations, setting to unity the parameters encoding the quasiparticle mass-renormalization weights~\cite{dmet-risb-1,dmet-risb-2}.
More recently, a more accurate extension of the GA, called "ghost Gutzwiller approximation" (g-GA) has been developed~\cite{Ghost-GA}, based on the idea of extending the GA variational space introducing auxiliary (ghost) fermionic degrees of freedom.

Here we present benchmark calculations of the Anderson lattice model (ALM) and demonstrate that, by construction, the GA cannot capture the interplay between Mott physics and the hybridization between correlated and itinerant degrees of freedom ---which generally coexist and whose interplay often governs the metal-insulator transition in real materials. 
We also show that, in some parameters regimes, this limitation of the GA can result in overestimating the Mott critical point  by orders of magnitude.
%
Finally, we demonstrate, both numerically and analytically, that the g-GA method resolves these problems, while remaining much less computationally demanding than DMFT.
Furthermore, we show that this method allows us to describe semi-analytically the spectral properties (both at low and high energies) throughout the entire phase diagram of the ALM, facilitating the physical interpretation of the numerical results.


\emph{Model}.--- 
We consider the ALM on a Bethe lattice, in the limit of infinite coordination number~\cite{GA-infinite-dim}:
\begin{align}
    \hat{H} &=  \sum_{<i,j>}\sum_{\sigma}  \left(t_{ij} + \delta_{ij}\epsilon_p\right)
   \pc_{i\sigma}\pa_{j\sigma} + \sum_{i}\frac{U}{2}\left(\hat{n}_{di}-1\right)^2 \nonumber\\
    &+ V\sum_{i\sigma}\left( \pc_{i\sigma}\da_{i\sigma} + \mathrm{H.c.}\right)
    - \mu\sum_i\hat{N}_i
    \nonumber
    \\
    &=
  \sum_{k\sigma}\sum_{\alpha\beta}[\tau_k]_{\alpha\beta}
  \,
  [\phi^\dagger_{k\sigma}]^\dagga_\alpha [\phi^\dagga_{k\sigma}]^\dagga_\beta
    +U\sum_i\dc_{i\uparrow}\da_{i\uparrow} \dc_{i\downarrow}\da_{i\downarrow}\,,
\end{align}
where:
$\pa_{i\sigma}$ and $\da_{i\sigma}$ are Fermionic annihilation operators, $\pc_{i\sigma}$ and $\dc_{i\sigma}$ are Fermionic creation operators, 
$i$ and $j$ are site labels, $\sigma$ is the spin, 
$<i,j>$ indicates that the corresponding summation is restricted to first nearest neighbours,
the hopping matrix $t_{ij}$ is uniform,
$\mu$ is the chemical potential,
$\hat{n}_{di}=\sum_{\sigma}\dc_{i\sigma}\da_{i\sigma}$,
$\hat{n}_{pi}=\sum_{\sigma}\pc_{i\sigma}\pa_{i\sigma}$,
$\hat{N}_i=\hat{n}_{di}+\hat{n}_{pi}$, 
$\phi^\dagger_{k\sigma}=(\pc_{k\sigma},\dc_{k\sigma})$,
$\phi^\dagger_{k\sigma}=\sum_i \mathcal{U}_{ik} \phi^\dagger_{i\sigma}$,
\begin{align}
    \tau_{k}=
    \begin{pmatrix}
    \epsilon_k+\epsilon_p-\mu & V  \\
    V & -U/2-\mu
    \end{pmatrix}
    \,,
\end{align}
the columns of $\mathcal{U}$ are the eigenvectors of $t$ and $\epsilon_k$ are its eigenvalues. 
From now on we fix the hopping matrix $t$ by using the half-bandwidth of $\epsilon_k$ (corresponding to a semi-circular density of states) as the energy unit.


\emph{Method}.--- 
Here we summarize the algorithmic structure of the g-GA and the GA, pointing out the key differences between these two methods, from a quantum-embedding perspective~\cite{Ghost-GA,Our-PRX,Lanata2016}.
For simplicity, below we focus on the ALM introduced above, while the general theory for arbitrary multi-orbital systems is summarized in the supplemental material~\cite{supplemental_material}.
For both the g-GA and the GA, the solution is obtained by calculating recursively the \emph{ground state} of 2 auxiliary systems: (1) the so-called "quasiparticle Hamiltonian" (QPH) and (2) the "embedding Hamiltonian" (EH).

The EH can be expressed in the following form:
\begin{align}
    \h_{\text{emb}}^{\text{g-GA}}&=
    \frac{U}{2}\left(\hat{n}_{d}-1\right)^2 -\mu\,\hat{n}_{d}
    +\sum_{\sigma}\sum_{a,b=1}^B\lambda^{\mathrm{c}}_{ab}\, \hat{f}^{\phantom{\dagger}}_{b\sigma}\hat{f}^{\dagger}_{a\sigma} 
    \nonumber\\
    &+\sum_{a=1}^B\sum_{\sigma}D_a \left( \hat{d}^{\dagger}_{\sigma}\hat{f}^{\phantom{\dagger}}_{a\sigma}+\mathrm{H.c.}\right) 
    \\
    \h_{\text{emb}}^{\text{GA}}&=
    \frac{U}{2}\left(\hat{n}_{d}-1\right)^2 -\mu\,\hat{n}_{d}
    + \sum_{\sigma}\lambda^c\, \hat{f}^{\phantom{\dagger}}_{\sigma}\hat{f}^{\dagger}_{\sigma}
    \nonumber\\
    &+\sum_{\sigma} D\left( \hat{d}^{\dagger}_{\sigma}\hat{f}^{\phantom{\dagger}}_{\sigma}+\mathrm{H.c.}\right)
    \,,
\end{align}
where the $\hat{d}$ operators correspond to the EH impurity degrees of freedom, $\hat{n}_{d}=\sum_{\sigma}\hat{d}^\dagger_{\sigma}\hat{d}^\dagga_{\sigma}$,
the $\hat{f}$ operators correspond to the EH bath degrees of freedom
and the parameters $D$ and $\lambda^c$ are determined self-consistently~\cite{supplemental_material}.
Note that, 
while in standard GA the bath of the EH has the same size of the impurity,
within the g-GA it contains a larger number of sites ($B>1$).
As in Refs.~\cite{Ghost-GA,exciton-mott-gga}, here we will set $B=3$
(the effect of increasing $B$, which would enlarge further the variational space, will be subject of future work).
After convergence, the expectation value of any local operator 
$\hat{O}[\dc_{i\alpha},\da_{i\alpha}]$ can be calculated from the
ground state $\ket{\Phi}$ of the EH as:
\be
\langle \hat{O} \rangle 
= \Av{\Phi}{\hat{O}[\hat{d}^{\dagger}_{\alpha},\hat{d}^{\dagga}_{\alpha}]}
\,.
\ee

The QPH can be expressed as:
\begin{align}
   \h_{\text{qp}}^{\text{g-GA}}&= \sum_{i\sigma}\sum_{a=1}^B l_a \fc_{ia\sigma}\fa_{ia\sigma} \nonumber\\
    &+ \sum_{i\sigma}\sum_{a=1}^B V\big(r_a \fc_{ia\sigma}\pa_{i\sigma} + \mathrm{H.c.}\big)\nonumber\\
    &+\sum_{i\sigma}\left(\epsilon_p-\mu\right) \pc_{i\sigma}\pa_{i\sigma}
    +\sum_{ij\sigma} t_{ij}\pc_{i\sigma}\pa_{j\sigma}
    \label{hqp-gga}
    \\
    \h_{\text{qp}}^{\text{GA}} &= \sum_{i\sigma}l\, \fc_{i\sigma}\fa_{i\sigma} + \sum_{i\sigma}V\big(r \fc_{i\sigma}\pa_{i\sigma} + \mathrm{H.c.}\big)\nonumber\\
    &+\sum_{i\sigma}\left(\epsilon_p-\mu\right) \pc_{i\sigma}\pa_{i\sigma}
    +\sum_{ij\sigma}t_{ij}\pc_{i\sigma}\pa_{j\sigma}
    \label{hqp-ga}
    \,,
\end{align}
where the $f_{ia\sigma}$ operators are "quasiparticle modes" residing in an auxiliary (enlarged) Hilbert space~\cite{supplemental_material}.
Once the parameters $l$ and $r$ are determined self-consistently in the form above~\cite{supplemental_material}, the resulting Green's function for the modes $\phi_{k\sigma}$ is:
\begin{align}
    \mathcal{G}(k,\omega)=\big[
    \omega-\tau_k-\Sigma(\omega)
    \big]^{-1}
    \,,
\end{align}
where the only non-zero entry of $\Sigma(\omega)$ is the $dd$ component, which is given by the following equations:
\begin{widetext}
\begin{align}
    \Sigma^{\text{g-GA}}_{dd}(\omega)&=
    \mu \! + \! \frac{U}{2} \! 
    + \! \frac{l_1}{r_1^2}
    \!-\! \omega\frac{1\!-\!r_1^2}{r_1^2}
    \!+\!\frac{(\omega\!-\!l_1)^2}{r_1^4}
    \big[
   (\omega\!-\!l_3)r_2^2+(\omega\!-\!l_2)r_3^2
    \big]
   \!\left[
    (\omega\!-\!l_2)(\omega\!-\!l_3)\!+\!\frac{\omega\!-\!l_1}{r_1^2}
    \big(
    r_2^2(\omega\!-\!l_3)\!+\!r_3^2(\omega\!-\!l_2)
    \big)
    \right]^{-1}
    \label{sigma-gga}
    \\
    \Sigma^{\text{GA}}_{dd}(\omega)&=
    \mu  + \frac{U}{2}
    + \frac{l}{r^2} - \omega\frac{1-r^2}{r^2}
    \label{sigma-ga}
    \,.
\end{align}
\end{widetext}

As demonstrated in the supplemental material~\cite{supplemental_material} (and shown in the calculations below): (i) the poles of
$\mathcal{G}(k,\omega)$ are located on top of the eigenvalues of the corresponding QPHs~\cite{hiddenQP-1,hiddenQP-2,hiddenQP-3,hiddenQP-4}, and
(ii) the resulting total spectral weight of the $d$ degrees of freedom is given by:
\begin{align}
    \int_{-\infty}^{\infty}
    \mathcal{A}^{\text{g-GA}}_{dd}(k,\omega)d\omega
    &=\sum_{a=1}^B r_a^2
    \label{totweight-gga}
    \\
    \int_{-\infty}^{\infty}
    \mathcal{A}^{\text{GA}}_{dd}(k,\omega)d\omega
    &=r^2
    \label{totweight-ga}
    \,,
\end{align}
where $\mathcal{A}^{\text{g-GA}}_{dd}$ and $\mathcal{A}^{\text{GA}}_{dd}$ are the g-GA and GA $d$-electron spectral functions, respectively, and $r^2$ is the GA $d$-electron quasiparticle weight~\cite{Gebhard-FL}.
From the parameters $R,\lambda$ and the ground state $\ket{\Psi_0}$ of the corresponding QPH,
it is also possible to calculate the expectation values of all non-local quadratic operators.
In particular:
\begin{align}
    \langle
    \dc_{i\sigma}\pa_{i\sigma}
    \rangle_{\text{g-GA}}
    &=\sum_{a=1}^B r_a 
    \Av{\Psi_0}{\fc_{ia\sigma}\pa_{i\sigma}}
    \label{hibr-gga}
    \\
    \langle
    \dc_{i\sigma}\pa_{i\sigma}
    \rangle_{\text{GA}}
    &=r 
    \Av{\Psi_0}{\fc_{i\sigma}\pa_{i\sigma}}
    \label{hibr-ga}
\,.
\end{align}

%
For completeness, the generalization of all analytical results listed above to arbitrary multi-orbital systems is given in the supplemental material~\cite{supplemental_material}.


\begin{figure} 
    \includegraphics[width=8.7cm]{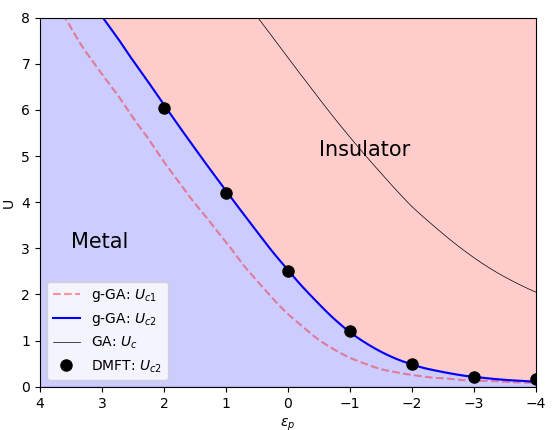}
    \caption{Paramagnetic
    phase diagram of the ALM on an infinite-coordination Bethe lattice, for $3$ electrons per site and $V=1$.
    The g-GA metal-insulator transition $U_{c2}$ and the end of the metal-insulator coexistence region $U_{c1}$ are marked in blue and red, respectively.
    The black dots are $U_{c2}$ values calculated with DMFT+CTQMC, at $T=0.01$.
    The gray line indicates the metal-insulator transition in bare GA.
    }
    \label{Figure1}
\end{figure}

\begin{figure*}[ht]
\begin{center}
\includegraphics[width=17.5cm]{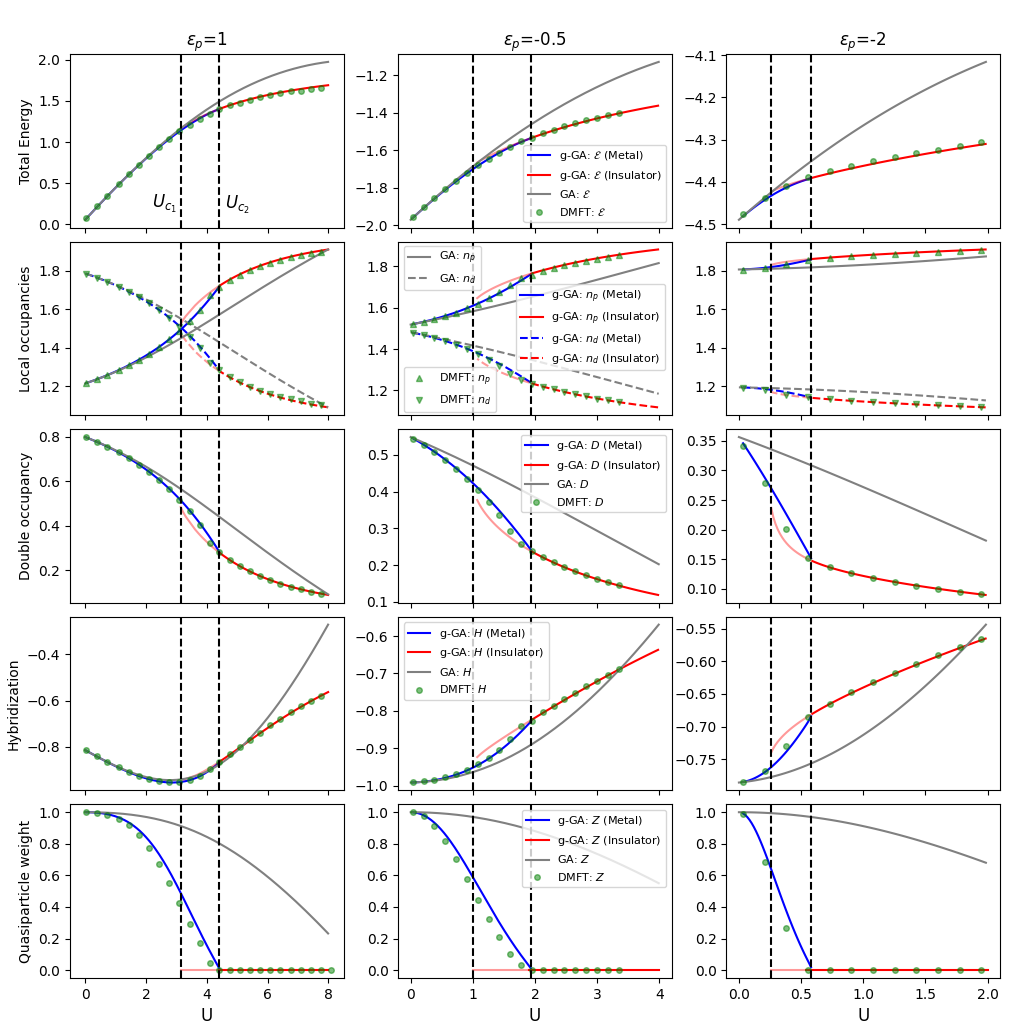}
\caption{
Behavior of the g-GA total energy $\mathcal{E}$, local occupancies $n_p=\langle\hat{n}_{pi}\rangle$ and $n_d=\langle\hat{n}_{di}\rangle$, $d$-electron double occupancy
$D=\langle\hat{n}_{di\uparrow}\hat{n}_{di \downarrow}\rangle$,
$p$-$d$ hybridization $H=\sum_\sigma\langle
    \dc_{i\sigma} p_{i\sigma}
    \rangle +\text{c.c.}$
and $d$-electron quasiparticle weight $Z$,
for the ALM on an infinite-coordination Bethe lattice, in comparison with DMFT and the bare GA.
The DMFT data are computed with CTQMC, at $T=0.01$ for $\epsilon_p=1$ and $\epsilon_p=-0.5$ and at $T=0.02$ for $\epsilon_p=-2$.
The vertical black dashed lines indicate $U_{c1}$ and $U_{c2}$.
}
\label{Figure2}
\end{center}
\end{figure*}

\emph{Results}.--- 
In Fig.~\ref{Figure1} we show the g-GA phase diagram of the ALM
(in the paramagnetic phase) for total occupancy $\langle \hat{N}_{i}\rangle=3$.
The g-GA results are compared with DMFT ---with the continuous time quantum Monte Carlo (CTQMC) impurity solver~\cite{ctqmc,ctqmc-Gull-RevModPhys,ctqmc-Haule}, at temperature $T=0.01$--- and with the bare GA.
%
Our benchmark calculations show that the g-GA phase diagram is consistent with previous work~\cite{ALM-1,ALM-2,ALM-3} and in remarkable agreement with DMFT.
As expected, both the g-GA method and the bare GA capture the fact that the Mott metal-insulator transition point $U_{c2}$ vanishes for $\eps_p\rightarrow -\infty$ ---corresponding to the limit where the $p$ degrees of freedom are gapped out.
However, the interaction $U_c$ of the metal-insulator transition 
is largely overestimated within the GA, especially for $\epsilon_p\ll -1$.
Note that the phase diagram for $\langle \hat{N}_{i}\rangle=1$ can be automatically inferred from our calculations above, as they are related to each other by a particle-hole transformation.

In Fig.~\ref{Figure2} we show the behavior of the g-GA total energy $\mathcal{E}$, the occupancies $n_p=\langle\hat{n}_{pi}\rangle$ and $n_d=\langle\hat{n}_{di}\rangle$, the $d$-electron double occupancy
$D=\langle\hat{n}_{di\uparrow}\hat{n}_{di \downarrow}\rangle$,
the $p$-$d$ "hybridization energy"
$H=\sum_\sigma\langle \dc_{i\sigma}\pa_{i\sigma} \rangle +\text{c.c.}$,
and the $d$-electron quasiparticle weight:
\begin{align}
    Z &= \left[ 1 - \frac{\partial \Sigma^{\mathrm{g-GA}}}{\partial\omega}\right]^{-1}_{\omega= 0}
    \nonumber
    \\
    &= \frac{\left(l_2 \, l_3 \, r_1^2 + l_1 \, l_3 \, r_2^2 + l_1 \, l_2 \, r_3^2\right)^2}
    {l_1^2 \, l_3^2 \, r_2^2 + l_2^2 \left(l_3^2 \, r_1^2 + l_1^2 \, r_3^2\right)}  
    \,.
\end{align}
The g-GA results are shown in comparison with the bare GA and DMFT.
While the GA solution is considered accurate only for weak interactions, we find that the agreement between g-GA and DMFT is remarkable for all observables, in all parameters regimes.

\begin{figure*}[ht]
\begin{center}
\includegraphics[width=17.5cm]{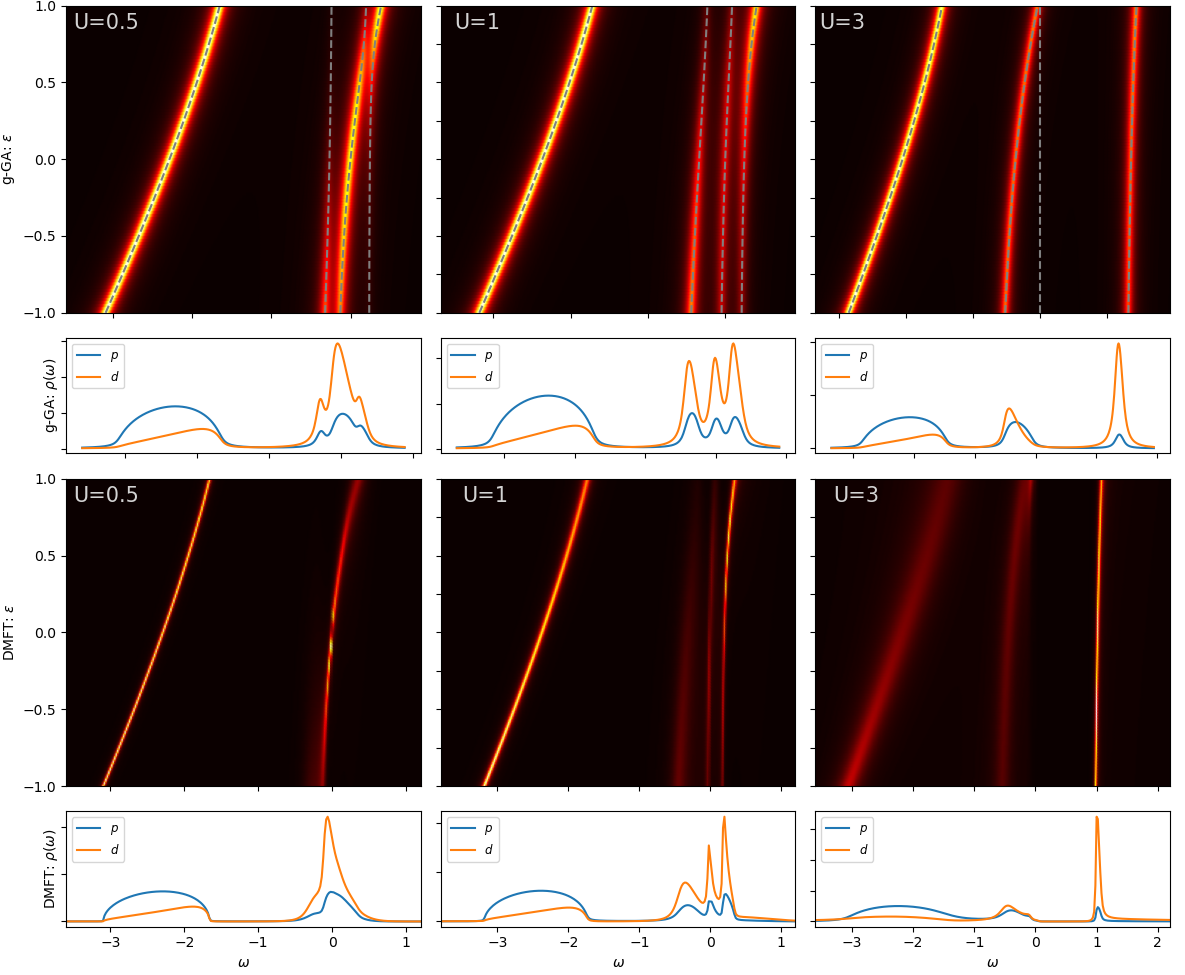}
\caption{
Energy-resolved spectral function and corresponding $p$ and $d$ local density of states $\rho(\omega)$, calculated
with g-GA (upper panels) and DMFT+CTQMC (lower panels), for $\epsilon_p=-1$ and 3 values of $U$.
The g-GA spectra is visualized using a small artificial smearing, $\Gamma=0.06$.
The g-GA quasiparticle bands are indicated by gray dashed lines. 
}
\label{Figure3}
\end{center}
\end{figure*}

Let us now analyze the single-particle Green's function.
In Fig.~\ref{Figure3} we consider $\epsilon_p=-1$ and 3 values of $U$,
showing the total g-GA 
energy-resolved spectral function:
\be
\mathcal{A}(k,\omega)=-\frac{1}{\pi}\text{Im}
\Tr[\mathcal{G}(k,\omega)]
\ee
and the $p$ and $d$ local density of states.
The DMFT spectra were obtained by performing analytical continuation with the maximum entropy method~\cite{JARRELL1996133}.
Interestingly, the g-GA captures systematically the main features of the DMFT spectra (including the Hubbard bands, and the hybridization between the $p$ and $d$ degrees of freedom).
Note that, to interpret the behavior of the g-GA spectra, it is possible to exploit its relation with the bands of the QPH [Eq.~\eqref{hqp-gga}], previously discussed in the methods section.
In particular, the relative position of the $p$ band with respect to the Fermi level is approximately encoded in the corresponding on-site energy
$\epsilon_p^*=\epsilon_p-\mu$, while the positions of the $d$-electron low-energy and high-energy excitations are approximately encoded in the variational parameters $l_a$ ($a=1,2,3$).
%


A key fact emerging from our benchmark calculations is that the GA can overestimate $U_c$ dramatically (especially for $\epsilon_p\ll -1$), see Fig.~\ref{Figure1}.
To explain this result we note that, by construction, the GA Mott transition occurs when the quasiparticle weight $Z=r^2$ vanishes, see Eq.~\eqref{sigma-ga}.
Therefore, the corresponding approximation to the Mott phase is such that $\langle\dc_{i\sigma}\pa_{i\sigma}\rangle_{\text{GA}}=0$, see Eq.~\eqref{hibr-ga}.
In other words, this method cannot describe simultaneously the Mott phase and the $p$-$d$ charge fluctuations.
But this is unrealistic for the ALM, where the $p$-$d$ hybridization effects are generally very large, not only in the weakly-interacting regime, but also for $U\simeq U_{c2}$ and $U>U_{c2}$, see Figs.~\ref{Figure2},\ref{Figure3}.
Because of the variational principle, this results in a systematic overestimation of the total energy as we approach the Mott phase (see Fig.~\ref{Figure2}), causing an overestimation of the metal-insulator transition point.
This point is documented in further detail in the supplemental material, where the GA overestimation of $U_{c}$ at $\epsilon_p\ll -1$ is explained in relation to a qualitative pathological behavior of the GA variational parameter $r$ in the narrow-bandwidth limit ($t\rightarrow 0)$.

Remarkably, since the g-GA captures the existence of the Hubbard bands, the right side of Eq.~\eqref{totweight-gga} never vanishes~\cite{Ghost-GA}.
Therefore, similar to DMFT, $\langle\dc_{i\sigma}\pa_{i\sigma}\rangle_{\text{g-GA}}$
remains finite even in the Mott phase (see Eq.~\eqref{hibr-gga}).
This shows that the ability of the g-GA of describing simultaneously the Mott physics and the $d$-$p$ hybridization is directly connected with its ability of describing the transfer of $d$-electron spectral weight to the Hubbard bands (which the bare GA lacks).


\emph{Conclusions}.--- 
We performed benchmark calculations of the ALM, showing that the g-GA provides us with results with accuracy comparable to DMFT, both for the ground-state and the spectral properties.
In particular, we showed that the g-GA is capable of describing accurately the interplay between the Mott physics and the hybridization between correlated and itinerant degrees of freedom, 
while the GA cannot describe simultaneously these effects.
This is particularly relevant for real-material calculations in combination with density functional theory~\cite{Anisimov_DMFT,Fang,Our-PRX,LDA+U}, where the correlated orbitals are generally very localized around their atomic positions~\cite{Haule10,PWF2} ---so that the interactions with their environment are mainly mediated by the itinerant modes (as in the ALM studied here).
In fact, it is well possible that the limitation of the GA here uncovered explains why, in some cases, simulating the properties of real materials with the GA requires to use unphysically-large Hubbard $U$~\cite{npj-lanata}, and suggests that multi-orbital implementations of the g-GA will resolve these problems.

From the computational standpoint, the g-GA is more expensive than the bare GA (as the bath of the EH contains additional degrees of freedom). On the other hand, its 
computational complexity remains much lower than DMFT.
In fact, the g-GA requires to calculate only the ground state of a finite-size impurity model (while in DMFT it is necessary to calculate the spectra of an impurity model with an infinite bath).
For example, in our DMFT calculations each CTQMC iteration (performed using $5\times 10^8$ Monte Carlo steps, in parallel, on 72 cores) required about 2 minutes of computational time. Instead, within our g-GA calculations each EH iteration (performed on a single core) required about 0.2-0.3 seconds.
Note also that the difference in computational complexity between DMFT and g-GA grows exponentially as a function of the impurity size.

A particularly promising perspective is the possibility of solving the g-GA equations with hybrid quantum-classical frameworks~\cite{YY-qc}, employing impurity solvers based on quantum algorithms such as variational quantum eigensolvers~\cite{mcclean2016theory, o2016scalable, kandala2017hardware, romero2018strategies}.
In fact, within the g-GA, realizing such program for real-material applications may require devices consisting of only tens of qubits, while it has been estimated that quantum computers with at least 100 logical qubits will be necessary for applications within DMFT~\cite{HybridQC}.
Furthermore, since the number of parameters characterizing the EH is finite, the recently-developed approach based on machine learning for the GA~\cite{Our-ML-actinides} will be applicable also to the g-GA, as we hope to show in future work.
Note also that the g-GA can be equivalently formulated in terms of the rotationally-invariant slave-boson mean-field theory~\cite{Ghost-GA}, which is based on an exact reformulation of the many-body problem. This line of interpretation may open the possibility of developing beyond-mean-field schemes, providing us with new routes for high-precision calculations.

\section*{Acknowledgements}
We thank Gabriel Kotliar for useful discussions.
We gratefully acknowledge funding from the Novo Nordisk Foundation through the Exploratory Interdisciplinary Synergy Programme project NNF19OC0057790.
We thank support from the VILLUM FONDEN through the Villum Experiment project 00028019 and the Centre of Excellence for Dirac Materials (Grant. No. 11744).
T.-H.L. was supported by the Computational Materials Sciences Program funded by the US Department of Energy, Office of Science, Basic
Energy Sciences, Materials Sciences and Engineering Division.
Work in Florida was supported by the NSF Grant No. 1822258, and the National High Magnetic Field Laboratory through the NSF Cooperative Agreement No. 1157490 and the State of Florida.


%

\end{document}


\newcommand{\cau}{\underline{c}^{\phantom{\dagger}}}
\newcommand{\ccu}{\underline{c}^\dagger}

\newcommand{\hu}{\underline{\hat{H}}}

\newcommand{\proj}{\hat{\mathcal{P}}_G}
\newcommand{\projR}{\hat{\mathcal{P}}_R}

\newcommand{\projRi}{\hat{\mathcal{P}}_{\mathbf{R}i}^{\phantom{\dagger}}}

\newcommand{\projRidagger}{\hat{\mathcal{P}}_{\mathbf{R}i}^{{\dagger}}}

\newcommand{\Hqp}{\hat{H}_{\text{qp}}}
\newcommand{\tlambda}{\tilde{\lambda}}
\newcommand{\tR}{\tilde{\mathcal{R}}}

\newcommand{\pa}{{p}^{\phantom{\dagger}}}
\newcommand{\pc}{{p}^\dagger}

\newcommand{\Dp}{\hat{\Delta}_{\text{p}}}
\newcommand{\Dh}{\hat{\Delta}_{\text{h}}}
\newcommand{\Deltap}{[\hat{\Delta}_{\text{p}}]}
\newcommand{\Deltah}{[\hat{\Delta}_{\text{h}}]}
\newcommand{\R}{\mathcal{R}}
\newcommand{\Rh}{\hat{\mathcal{R}}}

\newcommand{\uA}{|\underline{A},Ri\rangle}
\newcommand{\uB}{|\underline{B},R'j\rangle}

\newcommand{\E}{\mathcal{E}}
\newcommand{\G}{\mathcal{G}}
\newcommand{\Lag}{\mathcal{L}}
\newcommand{\M}{\mathcal{M}}
\newcommand{\N}{\mathcal{N}}
\newcommand{\U}{\mathcal{U}}
\newcommand{\F}{\mathcal{F}}
\newcommand{\V}{\mathcal{V}}
\newcommand{\C}{\mathcal{C}}
\newcommand{\I}{\mathcal{I}}
\newcommand{\s}{\sigma}
\newcommand{\up}{\uparrow}
\newcommand{\dw}{\downarrow}
\newcommand{\h}{\hat{H}}
\newcommand{\hilb}{\mathcal{H}}
\newcommand{\himp}{\hat{H}_{\text{imp}}}
\newcommand{\g}{\mathcal{G}^{-1}_0}
\newcommand{\D}{\mathcal{D}}
\newcommand{\A}{\mathcal{A}}
\newcommand{\projs}{\hat{\mathcal{S}}_d}
\newcommand{\K}{\textbf{k}}
\newcommand{\Q}{\textbf{q}}
\newcommand{\T}{\tau_{\ast}}
\newcommand{\io}{i\omega_n}
\newcommand{\eps}{\varepsilon}
\newcommand{\+}{\dag}
\newcommand{\su}{\uparrow}
\newcommand{\giu}{\downarrow}
\newcommand{\0}[1]{\textbf{#1}}
\newcommand{\ca}{c^{\phantom{\dagger}}}
\newcommand{\cc}{c^\dagger}
\newcommand{\aaa}{a^{\phantom{\dagger}}}
\newcommand{\aac}{a^\dagger}
\newcommand{\bba}{b^{\phantom{\dagger}}}
\newcommand{\bbc}{b^\dagger}
\newcommand{\da}{d^{\phantom{\dagger}}}
\newcommand{\dc}{d^\dagger}
\newcommand{\hfa}{\hat{f}^{\phantom{\dagger}}}
\newcommand{\hcc}{\hat{c}^\dagger}
\newcommand{\hca}{\hat{c}^{\phantom{\dagger}}}
\newcommand{\hfc}{\hat{f}^\dagger}
\newcommand{\fa}{f^{\phantom{\dagger}}}
\newcommand{\fc}{f^\dagger}
\newcommand{\ha}{h^{\phantom{\dagger}}}
\newcommand{\hc}{h^\dagger}
\newcommand{\be}{\begin{equation}}
\newcommand{\ee}{\end{equation}}
\newcommand{\bea}{\begin{eqnarray}}
\newcommand{\eea}{\end{eqnarray}}
\newcommand{\ba}{\begin{eqnarray*}}
\newcommand{\ea}{\end{eqnarray*}}
\newcommand{\dagga}{{\phantom{\dagger}}}
\newcommand{\bR}{\mathbf{R}}
\newcommand{\bQ}{\mathbf{Q}}
\newcommand{\bq}{\mathbf{q}}
\newcommand{\bqp}{\mathbf{q'}}
\newcommand{\bk}{\mathbf{k}}
\newcommand{\bh}{\mathbf{h}}
\newcommand{\bkp}{\mathbf{k'}}
\newcommand{\bp}{\mathbf{p}}
\newcommand{\bL}{\mathbf{L}}
\newcommand{\bRp}{\mathbf{R'}}
\newcommand{\bx}{\mathbf{x}}
\newcommand{\by}{\mathbf{y}}
\newcommand{\bz}{\mathbf{z}}
\newcommand{\br}{\mathbf{r}}
\newcommand{\Ima}{{\Im m}}
\newcommand{\Rea}{{\Re e}}
\newcommand{\Pj}[2]{|#1\rangle\langle #2|}
\newcommand{\ket}[1]{\vert#1\rangle}
\newcommand{\bra}[1]{\langle#1\vert}
\newcommand{\setof}[1]{\left\{#1\right\}}
\newcommand{\fract}[2]{\frac{\displaystyle #1}{\displaystyle #2}}
\newcommand{\Av}[2]{\langle #1|\,#2\,|#1\rangle}
\newcommand{\av}[1]{\langle #1 \rangle}
\newcommand{\Mel}[3]{\langle #1|#2\,|#3\rangle}
\newcommand{\Avs}[1]{\langle \,#1\,\rangle_0}
\newcommand{\eqn}[1]{(\ref{#1})}
\newcommand{\Tr}{\mathrm{Tr}}

\newcommand{\Vb}{\bar{\mathcal{V}}}
\newcommand{\Vd}{\Delta\mathcal{V}}
\def\P{P_{02}}
\newcommand{\Pb}{\bar{P}_{02}}
\newcommand{\Pd}{\Delta P_{02}}
\def\t{\theta_{02}}
\newcommand{\tb}{\bar{\theta}_{02}}
\newcommand{\td}{\Delta \theta_{02}}
\newcommand{\Rb}{\bar{R}}
\newcommand{\Rd}{\Delta R}

\title{Supplemental Material:\\
	Quantum-embedding description of the Anderson lattice model with \\ the ghost Gutzwiller Approximation}

\author{Marius S. Frank}
\affiliation{Department of Physics and Astronomy, Aarhus University, 8000, Aarhus C, Denmark}
\author{Tsung-Han Lee}
\affiliation{Physics and Astronomy Department, Rutgers University, Piscataway, New Jersey 08854, USA}
\author{Gargee Bhattacharyya}
\affiliation{Department of Physics and Astronomy, Aarhus University, 8000, Aarhus C, Denmark}
\author{Pak Ki Henry Tsang}
\affiliation{Department of Physics and National High Magnetic Field Laboratory, Florida State University, Tallahassee, Florida 32306, USA}
\author{Victor Quito}
\affiliation{Department of Physics and Astronomy, Iowa State University, Ames, Iowa 50011, USA}
\affiliation{Department of Physics and National High Magnetic Field Laboratory, Florida State University, Tallahassee, Florida 32306, USA}
\author{Vladimir Dobrosavljevi\'c}
\affiliation{Department of Physics and National High Magnetic Field Laboratory, Florida State University, Tallahassee, Florida 32306, USA}
\author{Ove Christiansen}
\affiliation{Department of Chemistry, Aarhus University, 8000, Aarhus C, Denmark}
\author{Nicola Lanat\`a}
\altaffiliation{Corresponding author: lanata@phys.au.dk}
\affiliation{Department of Physics and Astronomy, Aarhus University, 8000, Aarhus C, Denmark}
\affiliation{Nordita, KTH Royal Institute of Technology and Stockholm University, Roslagstullsbacken 23, 10691 Stockholm, Sweden}

\begin{abstract}

\end{abstract}

\maketitle

\section{The $\text{g}$-GA theory}

For completeness, here we provide a comprehensive derivation of the g-GA method, presenting the theory from a slightly different perspective with respect to Ref.~\cite{Ghost-GA} (more direct and concise, but equivalent).

\subsection{The Hamiltonian}

We consider a generic multi-orbital Fermionic Hamiltonian represented as follows:
\begin{align}
\h &=
\sum_{\bR \bRp} \sum_{ij}
\sum_{\alpha=1}^{{\nu_i}} 
\sum_{\beta=1}^{{\nu_j}} 
{t}_{\bR i,\bRp j}^{\alpha\beta}\,
\cc_{\bR i\alpha}\ca_{\bRp j\beta}+
\sum_{\bR}\sum_{i\geq 1}
\h_{\bR i}^{\text{loc}}[\cc_{\bR i\alpha},\ca_{\bR i\alpha}]
\nonumber
\\
&=\sum_{\bk} \sum_{ij}
\sum_{\alpha=1}^{{\nu_i}} 
\sum_{\beta=1}^{{\nu_j}} 
{t}_{\bk,ij}^{\alpha\beta}\,
\cc_{\bk i\alpha}\ca_{\bk j\beta}+
\sum_{\bR}\sum_{i\geq 1}
\h_{\bR i}^{\text{loc}}[\cc_{\bR i\alpha},\ca_{\bR i\alpha}]
\,,
\label{hubb}
\end{align}
where $\bR$ indicates the unit-cell label, $\bk$ is the corresponding crystal-momentum, while $i$ and $\alpha$ classify the Fermionic degrees of freedom (both spin and orbital) according to the following convention: (1) all modes with $i=0$ are "uncorrelated", i.e., they appear only in the quadratic part of $\h$ (as the $p$ modes of the ALM, introduced in the main text); (2) all modes with $i\geq 1$ are "correlated" (as the $d$ modes of the ALM), i.e., $\h_{\bR i}^{\text{loc}}$ are generic operators (including interaction terms) constructed from $\cc_{\bR i\alpha}\ca_{\bR j\beta}$.
As in the multi-orbital GA, we assume that:
\be
{t}_{\bR i, \bR i}^{\alpha_i\beta_i}=0\quad\forall\,i\geq 1
\,,
\ee
and that $\h_{\bR i}^{\text{loc}}$ includes both the two-body local terms of the Hamiltonian and the one-body local term represented as:
\be
\h^{(1)}_{\text{loc}} \,=\,
\sum_{\bR} \sum_{i\geq 1}
\sum_{\alpha,\beta=1}^{\nu_i} 
[{t}^{\text{loc}}_{i}]_{\alpha\beta}\,
\cc_{\bR i\alpha}\ca_{\bR i\beta}
=
\sum_{\bk} \sum_{i\geq 1}
\sum_{\alpha,\beta=1}^{\nu_i} 
[{t}^{\text{loc}}_{i}]_{\alpha\beta}\,
\cc_{\bk i\alpha}\ca_{\bk i\beta}
\,.
\label{hloc-quadratic}
\ee

Note that in first-principle calculations of real materials based on density functional theory in combination with many-body techniques~\cite{Anisimov_DMFT,Fang,Our-PRX,LDA+U}, such as the GA or DMFT, the solution is obtained by solving recursively multi-orbital Hamiltonian with the structure of Eq.~\eqref{hubb}~\cite{Our-PRX}.
It is important to note that, within this context, the correlated degrees of freedom correspond to localized orbitals~\cite{Haule10,PWF2}. Therefore, as pointed out in the conclusions of the main text, the hopping parameters between correlated and uncorrelated modes is much larger with respect to the hopping between correlated modes (as in the ALM studied in the main text). 
Therefore, the pathology of the GA method uncovered in this work and our observation that the g-GA extension resolves this problem are very relevant in ab-initio calculations of correlated materials.

\begin{figure} 
    \includegraphics[width=7.7cm]{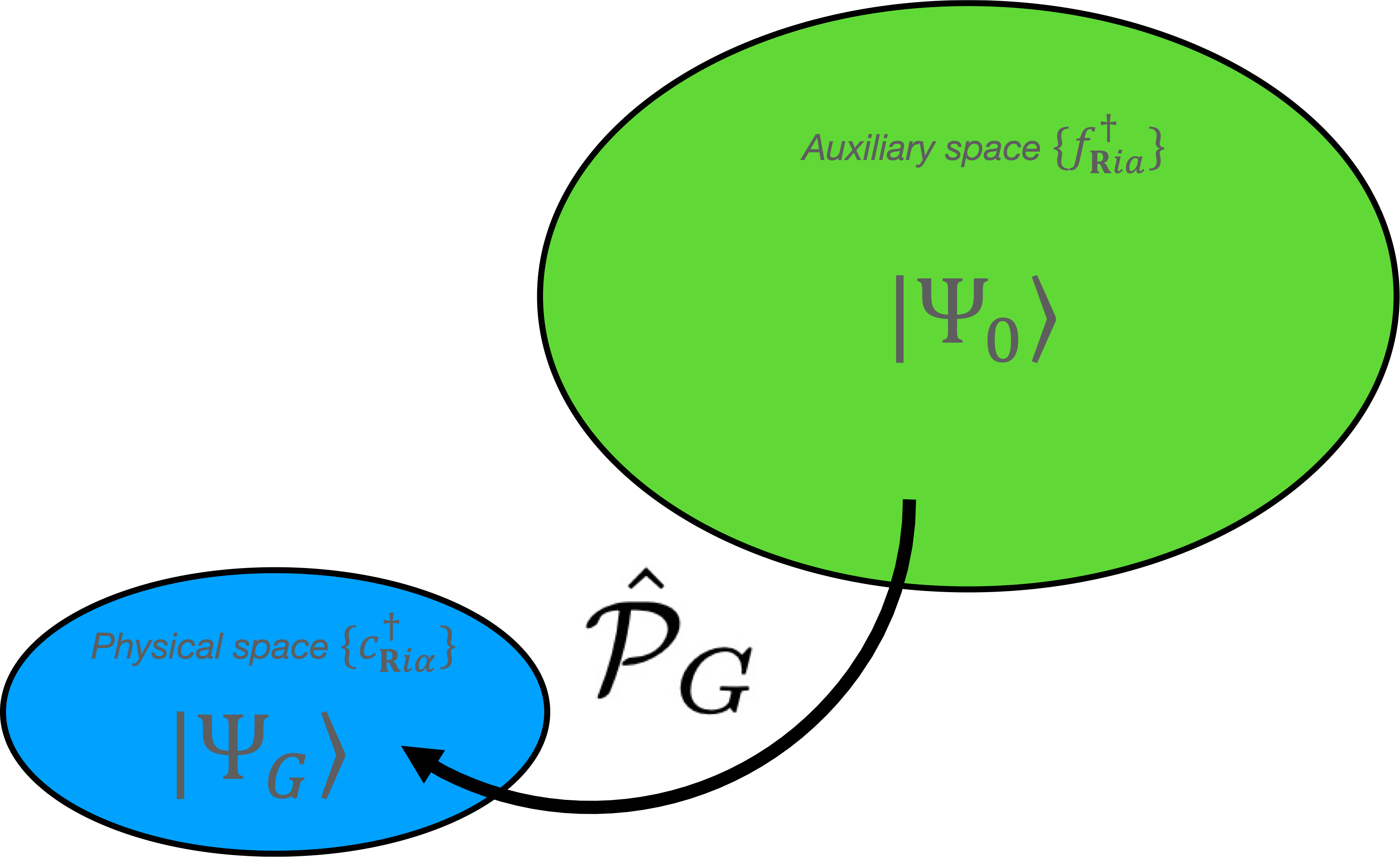}
    \caption{Schematic representation of the g-GA variational ansatz. The wavefunction $\ket{\Psi_G}$ is constructed by mapping a generic single-particle wavefunction $\ket{\Psi_0}$ constructed in an auxiliary Hilbert space, using the operator $\proj$.
    Both $\ket{\Psi_0}$ and $\proj$ are determined variationally}
    \label{FigureS1}
\end{figure}

\subsection{The $\text{g}$-GA variational ansatz}

The g-GA consists in minimizing the expectation value of $\h$
with respect to a wave function represented as follows:
\begin{align}
    \ket{\Psi_G}&=\proj\ket{\Psi_0}
\label{GWF}
\\
\proj&=\prod_{i\geq 1}\projRi
\,,
\end{align}
where $\ket{\Psi_0}$ is a single-particle wavefunction constructed in an auxiliary Hilbert space, generated by $\tilde{\nu}_i>\nu_i$ degrees of freedom $\fc_{\bR ia}$ $(a=1,..,\tilde{\nu}_i)$ for each $(\bR,i)$, while:
\begin{align}
\projRi&=\sum_{\Gamma=0}^{2^{\nu_i}-1}
\,
\sum_{n=0}^{2^{\tilde{\nu}_i}-1}
[\Lambda_{i}]_{\Gamma n}
\ket{\Gamma,{\bR i}}\bra{n,{\bR i}}
\\
\ket{\Gamma,{\bR i}}&=[\cc_{\bR i1}]^{q_1(\Gamma)} ...
[\cc_{\bR i q_{\nu_i}}]^{q_{\nu_i}(\Gamma)}\,\ket{0}
\\
\ket{n,{\bR i}}&=[\fc_{\bR i1}]^{q_1(n)} ...
[\fc_{\bR i q_{\tilde{\nu}_i}}]^{q_{\tilde{\nu}_i}(n)}\,\ket{0}
\end{align}
is an operator mapping the local auxiliary-space states into the physical space, see Fig.~\ref{FigureS1}, where $q_i(j)$ represents the $i$-th digit of $j$ in binary representation, and the matrix $\Lambda_i$ controls how $\projRi$
modifies the weight of the local electronic configurations.
The key reason why the g-GA variational ansatz generalizes the standard multi-orbital GA theory is that $\tilde{\nu}_i>\nu_i$.

In the present work we focus on the normal phase. Note that, to enforce the condition that $\ket{\Psi_G}$ is an eigenstate of the number operator, we do \emph{not}
need to assume that $\projRi$ commutes with the number operator, but only that:
\be
\sum_{j=1}^{\tilde{\nu}_i} q_j(n)-
\sum_{j=1}^{\nu_i} q_j(\Gamma)=m_i\quad
\forall\,\Gamma,n \, | \,
[\Lambda_i]_{\Gamma n}\neq 0\,,
\label{mi}
\ee
where $m_i$ is integer.
Our choice of $m_i$ (that, in principle, could be considered as an arbitrary variational parameter) will be specified below.

As in the standard multi-orbital GA, the variational wave function is restricted by the following
conditions:
\begin{align}
\Av{\Psi_0}{\projRidagger\projRi} &= \langle\Psi_0|\Psi_0\rangle=1
\label{gconstr1}
\\
\Av{\Psi_0}{\projRidagger\projRi\,\fc_{\bR ia}\fa_{\bR ib}} &=
\Av{\Psi_0}{\fc_{\bR ia}\fa_{\bR ib}}\qquad\forall\,a,b=1,...,\tilde{\nu}_i 
\,,
\label{gconstr2}
\end{align}
which are commonly called "Gutzwiller constraints".
Furthermore, the so-called "Gutzwiller Approximation", which becomes
exact in the limit of infinite coordination number~\cite{Gutzwiller3,GA-infinite-dim}
---where Dynamical Mean Field Theory (DMFT) is exact~\cite{DMFT}--- is assumed.

\subsection{The $\text{g}$-GA variational-energy components}

As in the standard multi-orbital GA, our goal is to evaluate the variational energy:
\be
\mathcal{E}(\Psi_0,\{\Lambda_i\})=\Av{\Psi_0}{\proj^\dagger\hat{H}\proj}\,.
\label{varen}
\ee
The only difference is that the operator $\proj^\dagger\hat{H}\proj$ and the operators appearing in Eqs.~\eqref{gconstr1}-\eqref{gconstr2}  reside within the auxiliary (extended) Hilbert space generated by the $\fc_{Ria}$ operators, where $a\in\{1,..,\tilde{\nu}_i\}$ and $\tilde{\nu}_i\geq \nu_i$.
In fact, if we set $\tilde{\nu}_i = \nu_i$ we recover exactly the original multi-orbital GA theory.

As we explain below for completeness, all of the formal steps leading to the quantum-embedding formulation of the GA can be essentially repeated for the g-GA following Ref.~\cite{Our-PRX}, even if $\tilde{\nu}_i > \nu_i$.
In particular, by employing the Gutzwiller approximation~\cite{Gutzwiller3,GA-infinite-dim}
and assuming the Gutzwiller constraints [Eqs.~\eqref{gconstr1},\eqref{gconstr2}], it can be shown that:
\begin{align}
\Av{\Psi_0}{\proj^\dagger\,\cc_{\bk i\alpha}\ca_{\bk j\beta}\,\proj}
&=\sum_{a=1}^{\tilde{\nu}_i}\sum_{b=1}^{\tilde{\nu}_j}
\Av{\Psi_0}{\left([\mathcal{R}_i]_{a\alpha}\fc_{\mathbf{R}ia}\right)\left([\mathcal{R}_j]^\dagger_{\beta b}\fa_{\mathbf{R}'jb}\right)}
\qquad\forall\,(\mathbf{R},i)\neq (\mathbf{R}',j)\,,
\label{hoppav}
\\
\Av{\Psi_0}{\proj^\dagger\,\h_{\bR i}^{\text{loc}}[\cc_{\bR i\alpha},\ca_{\bR i\alpha}]\,\proj}
&=\Av{\Psi_0}{\projRidagger\,\h_{\bR i}^{\text{loc}}[\cc_{\bR i\alpha},\ca_{\bR i\alpha}]\,\projRi}\,,
\label{locav}
\end{align}
where the $\tilde{\nu}_i\times \nu_i$ matrices $\mathcal{R}_i$ are the solution of the following linear equation:
\be
\Av{\Psi_0}{\projRidagger \cc_{\bR i\alpha}\projRi \fa_{\bR ia}}
=\sum_{b=1}^{\tilde{\nu}_i}[\mathcal{R}_i]_{b\alpha}
\Av{\Psi_0}{\fc_{\bR i b} \fa_{\bR ia}}\,.
\label{Rdef}
\ee

\subsection{Local $\ket{\Psi_0}$ averages as a function of variational parameters}

In the previous subsection we reduced the problem of evaluating the total-energy components to Eqs.~\eqref{hoppav}-\eqref{Rdef}.
Since all of these equations, as well as the Gutzwiller constraints [Eqs.~\eqref{gconstr1},\eqref{gconstr2}], involve expectation values with respect to $\ket{\Psi_0}$ of "local operators" (i.e., involving only $\fc_{\bR ia}$ and $\fa_{\bR ia}$ degrees of freedom at fixed $(\bR,i)$),
it is useful to introduce the corresponding local reduced density matrix of $\ket{\Psi_0}$.

By exploiting the fact that $\ket{\Psi_0}$ is a single-particle wavefunction (i.e., Wick's theorem applies to it), it can be readily verified that its reduced density matrix to the $(\bR,i)$ subsystem is given by:
\be
\hat{P}^0_{\bR i}\propto \exp\left\{-\sum_{a,b=1}^{\tilde{\nu}_i} \left[\ln\left(\frac{1-{^t\!\Delta_i}}{^t\!\Delta_i}\right)\right]_{ab}
\fc_{\bR i a} \fa_{\bR ib}
\right\}\,,
\label{hatp0}
\ee
where the entries of the $\tilde{\nu}_i\times \tilde{\nu}_i$ matrix $\Delta_i$ are given by:
\be
[\Delta_i]_{ab}=\Av{\Psi_0}{\fc_{\bR i a} \fa_{\bR ib}}
\ee
and ${^t\!\Delta_i}$ indicates the transpose of $\Delta_i$.

From the definitions above, it can be straightforwardly verified that:
\begin{align}
    \Av{\Psi_0}{\projRidagger\projRi}&=
    \Tr\big[P^0_i\Lambda^\dagger_i\Lambda^\dagga_i
    \big]
    \label{loc1}
    \\
    \Av{\Psi_0}{\projRidagger\projRi\,\fc_{\bR ia}\fa_{\bR ib}}&=
    \Tr\big[P^0_i\Lambda^\dagger_i\Lambda^\dagga_i
    \tilde{F}_{ia}^\dagger \tilde{F}_{ib}^\dagga
    \big]
    \\
    \Av{\Psi_0}{\projRidagger\,\h_{\bR i}^{\text{loc}}[\cc_{\bR i\alpha},\ca_{\bR i\alpha}]\,\projRi}&=
    \Tr\big[P^0_i\Lambda^\dagger_i
    \h_{\bR i}^{\text{loc}}[F^\dagger_{i\alpha},F^\dagga_{i\alpha}]
    \Lambda^\dagga_i\big]
    \\
    \Av{\Psi_0}{\projRidagger \cc_{\bR i\alpha}\projRi \fa_{\bR ia}}&=
    \Tr\big[P^0_i\Lambda^\dagger_i
    F_{i\alpha}^\dagger \Lambda^\dagga_i \tilde{F}_{ib}^\dagga\big]
    \label{loc4}
    \,,
\end{align}
where $\Tr$ is the trace and:
\begin{align}
    [{P}^0_{i}]_{nn'}&=\langle n,\mathbf{R}i |
    \hat{P}^0_{\bR i}
    | n', \mathbf{R}i\rangle
    \qquad (n,n'\in\{0,..,2^{\tilde{\nu}_i}-1\})
    \\
    [F_{i\alpha}]_{\Gamma\Gamma'}&=
    \langle \Gamma,\mathbf{R}i |
    \ca_{\bR i\alpha}
    | \Gamma', \mathbf{R}i\rangle
    \qquad (\Gamma,\Gamma'\in\{0,..,2^{{\nu}_i}-1\})
    \label{FGamma}
    \\
    [\tilde{F}_{ia}]_{nn'}&=
    \langle n,\mathbf{R}i |
    \fa_{\bR i a}
    | n', \mathbf{R}i\rangle
    \qquad (n,n'\in\{0,..,2^{\tilde{\nu}_i}-1\})
    \label{Fn}
    \,.
\end{align}

\subsection{Matrix of slave-boson amplitudes}

Following Refs.~\cite{lanata-barone-fabrizio,equivalence_GA-SB,Our-PRX}, 
we introduce the so-called matrix of slave-boson (SB)
amplitudes~\cite{Fresard1992,Georges,Lanata2016}:
\be
\phi_i=\Lambda_i\sqrt{P^0_i}\,.
\label{sba}
\ee

By substituting Eq.~\eqref{sba} in Eqs.~\eqref{loc1}-\eqref{loc4}, it can be readily verified that the Gutzwiller constraints can be rewritten as follows:
\begin{align}
\Tr\big[\phi^\dagger_i\phi^\dagga_i\big] &= \langle\Psi_0|\Psi_0\rangle=1
\label{gconstr1sb}
\\
\Tr\big[\phi^\dagger_i\phi^\dagga_i \tilde{F}_{ia}^\dagger \tilde{F}_{ib}^\dagga\big]
    &=
\Av{\Psi_0}{\fc_{\bR ia}\fa_{\bR ib}}=[\Delta_i]_{ab}
\qquad\forall\,a,b=1,...,\tilde{\nu}_i 
\label{gconstr2sb}
\end{align}
and that Eq.~\eqref{Rdef} can be rewritten as follows:
\be
\Tr\big[\phi^\dagger_i
    F_{i\alpha}^\dagger \phi^\dagga_i \tilde{F}_{ia}^\dagga\big]
=\sum_{c=1}^{\tilde{\nu}_i}[\mathcal{R}_i]_{c\alpha}\,
[\Delta_i(1-\Delta_i)]^{\frac{1}{2}}_{ca}
\,.
\label{Rdefsb}
\ee

\subsection{Embedding mapping}

Following Ref.~\cite{Our-PRX}, we introduce the so called "embedding states," which are related to the SB amplitudes as follows:
\begin{align}
    \ket{\Phi_i}=\sum_{\Gamma=0}^{2^{\nu_i}-1} \,\sum_{n=0}^{2^{\tilde{\nu}_i}-1}
    e^{i\frac{\pi}{2}N(n)(N(n)-1)}
    \,[\phi_i]_{\Gamma n} \,
    \ket{\Gamma;i}\otimes U_{\text{PH}}\ket{n;i}
    \label{ehstates}
\end{align}
where
\begin{align}
    \ket{\Gamma;i}&= [\hat{c}^\dagger_{i1}]^{q_1(\Gamma)} ...
[\hat{c}^\dagger_{i q_{\tilde{\nu}_i}}]^{q_{{\nu}_i}(\Gamma)}\,\ket{0}
    \\
    \ket{n;i}&= [\hat{f}^\dagger_{i1}]^{q_1(n)} ...
[\hat{f}^\dagger_{i q_{\tilde{\nu}_i}}]^{q_{\tilde{\nu}_i}(n)}\,\ket{0}
    \,,
\end{align}
$U_{\text{PH}}$ is a particle-hole transformation acting over the $\ket{n;i}$ states and
\be
N(n)=\sum_{a=1}^{\tilde{\nu}_i} q_a(n)
\,.
\ee

Note that the set of all embedding states represented as in Eq.~\eqref{ehstates} constitute a Fock space, corresponding to an "impurity" (generated by the Fermionic degrees of freedom $\hat{c}_{i\alpha}$, $\alpha\in\{1,..,\nu_i\}$) and a "bath" (generated by the Fermionic degrees of freedom $\hat{f}_{i a}$, $a\in\{1,..,\tilde{\nu}_i\}$).

\begin{figure} 
    \includegraphics[width=7.7cm]{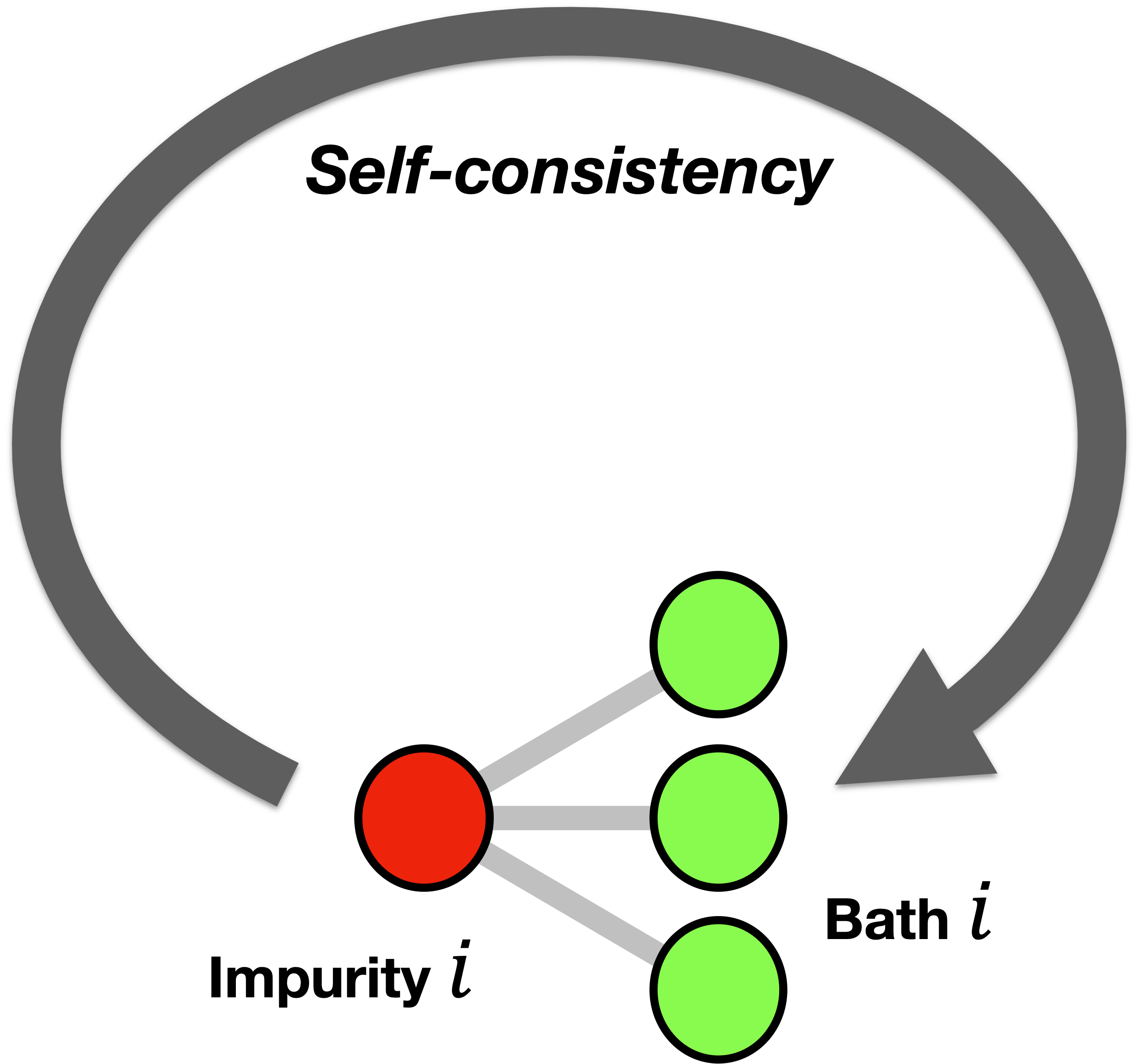}
    \caption{Schematic representation of the quantum-embedding algorithmic structure of the g-GA method. Here the number of bath sites $B=\tilde{\nu}_i/\nu_i$ (green circles) in the embedding Hamiltonian is set to $3$ (as in the calculations of the main text).
    }
    \label{FigureS2}
\end{figure}

The case $B=1$ corresponds to the standard GA theory, where the bath has the same size of the impurity.
In the present work we assumed that $\tilde{\nu}_i=B\nu_i$ and set $B=3$, see Fig.~\ref{FigureS2}.
Furthermore, we set $m_i=(\tilde{\nu}_i-\nu_i)/2=2$ (see Eq.~\eqref{mi}).
Note that, from the definitions [Eqs.~\eqref{sba},\eqref{ehstates}], it follows that this is equivalent to assume that 
$\ket{\Phi_i}$ has a total of $(\tilde{\nu}_i+\nu_i)/2=$4 electrons (i.e., that the embedding states are half-filled).

It can be readily verified by inspection that the Gutzwiller constraints can be rewritten as follows:
\begin{align}
\langle\Phi_i|\Phi_i\rangle &= \langle\Psi_0|\Psi_0\rangle=1
\label{gconstr1eh}
\\
\Av{\Phi_i}{\hat{f}^\dagga_{i b}\hat{f}^\dagger_{i a}}
&=
\Av{\Psi_0}{\fc_{\bR ia}\fa_{\bR ib}}=[\Delta_i]_{ab}
\qquad\forall\,a,b=1,...,\tilde{\nu}_i 
,,
\label{gconstr2eh}
\end{align}
the expectation value of the local terms of $\hat{H}$ can be calculated as:
\be
\Av{\Psi_0}{\projRidagger\,\h_{\bR i}^{\text{loc}}[\cc_{\bR i\alpha},\ca_{\bR i\alpha}]\,\projRi} =
\Av{\Phi_i}{\h_{\bR i}^{\text{loc}}[\hat{c}^\dagger_{ i\alpha},\hat{c}^\dagga_{i\alpha}]}
\ee
and that Eq.~\eqref{Rdef} can be rewritten as:
\be
\Av{\Psi_0}{\hat{c}^\dagger_{i \alpha}\hat{f}^\dagger_{i a}}
=\sum_{c=1}^{\tilde{\nu}_i}[\mathcal{R}_i]_{c\alpha}\,
[\Delta_i(1-\Delta_i)]^{\frac{1}{2}}_{ca}
\,.
\label{Rdefeh}
\ee

In summary, by substituting Eqs.~\eqref{hoppav} and \eqref{locav} in Eq.~\eqref{varen}
and using the equations above,
we deduce that the goal is to minimize the following expression for the variational energy:
\be
\mathcal{E}=
\sum_{\bR\bR'}\sum_{i,j\geq 0}\sum_{a,b=1}^{\bar{\nu}_i}\left[\R_i^\dagga t_{\bR i,\bR'j}\R^{\dagger}_j\right]_{ab} \fc_{\bR ia}\fa_{\bR' jb}
+
\sum_{\bR}\sum_{i\geq 1}
\Av{\Phi_i}{\h_{\bR i}^{\text{loc}}[\hat{c}^\dagger_{ i\alpha},\hat{c}^\dagga_{i\alpha}]}
\ee
while fulfilling the Gutzwiller constraints [Eqs.~\eqref{gconstr1eh}, \eqref{gconstr2eh}].

\subsection{Lagrange formulation of the g-GA} \label{covariance-sec}

Following Refs.~\onlinecite{Our-PRX,Lanata2016,Ghost-GA}, the constrained energy-minimization problem described above can be cast in terms of the following Lagrange function:

\bea
&&\Lag[
{\Phi},E^c;\,  \R,\lambda;\, \D, \lambda^{c};\,\Delta, \Psi_0, E]=
\frac{1}{\mathcal{N}}\Av{\Psi_0}{\h_{\text{qp}}[\R,\lambda]}
+E\!\left(1\!-\!\langle\Psi_0|\Psi_0\rangle\right)
\nonumber\\&&\quad\quad
+\sum_{i\geq 1}\left[\Av{\Phi_i}{\h_i^{\text{emb}}[\D_i,\lambda_i^c]}
+E^c_i\!\left(1-\langle \Phi_i | \Phi_i \rangle
\right)\right]
\nonumber\\&&\quad\quad
-\sum_{i\ge 1}\left[
\sum_{a,b=1}^{\tilde{\nu}_i}\big(
\left[\lambda_i\right]_{ab}+\left[\lambda^c_i\right]_{ab}\big)\left[\Delta_{i}\right]_{ab}
+\sum_{c,a=1}^{\tilde{\nu}_i}\sum_{\alpha=1}^{\nu_i}
\big(
\left[\D_{i}\right]_{a\alpha}\left[\R_{i}\right]_{c\alpha}
\left[\Delta_{i}(1-\Delta_{i})\right]^{\frac{1}{2}}_{ca}
+\text{c.c.}\big)
\right]\,,
\label{Lag-SB-emb}
\eea
where $\mathcal{N}$ is the total number of unit cells,
$E$ and $E^c$ are real numbers, $\Delta_i$, $\lambda^c_i$ and $\lambda_i$ are
$\tilde{\nu}_i\times \tilde{\nu}_i$ Hermitian matrices, $\D_i$ and $\R_i$ are generic (rectangular) $\tilde{\nu}_i\times {\nu}_i$ matrices.
The auxiliary Hamiltonians $\Hqp$ and $\h_{\text{emb}}$,
which are called "quasiparticle Hamiltonian" and "Embedding Hamiltonian,"
respectively, are defined as follows:
\begin{align}
    \h_{\text{qp}}&=\sum_{\bR\bR'}\sum_{i,j\geq 0}\sum_{a,b=1}^{\bar{\nu}_i}\left[\R_i^\dagga t_{\bR i,\bR'j}\R^{\dagger}_j\right]_{ab}
    \fc_{\bR ia}\fa_{\bR' jb}
    + \sum_{\bR}\sum_{i\geq 0}\sum_{a,b=1}^{\bar{\nu}_i} \left[ \lambda_i \right]_{ab}\fc_{\bR ia} \fa_{\bR ib}
    \label{hqp}
    \\
    \h_i^{\text{emb}}&= \hat{H}^{\mathrm{loc}}_{i}\big[\{\hat{c}^{\dagger}_{ i},\hat{c}^{\phantom{\dagger}}_{i}\} \big] + \sum_{a=1}^{\tilde{\nu}_i}\sum_{\alpha=1}^{\nu_i}\left[\D_i\right]_{a\alpha}\hat{c}^{\dagger}_{i\alpha}\hat{f}^{\phantom{\dagger}}_{ia}
   +\sum_{a,b=1}^{\tilde{\nu}_i}\left[\lambda^c_i\right]_{ab}\hat{f}^{\phantom{\dagger}}_{ib}\hat{f}^{\dagger}_{ia}
   \quad\forall\,i\geq 1
   \,.
   \label{hemb}
\end{align}

The saddle-point of the Lagrangian $\Lag$ defined in Eq. \eqref{Lag-SB-emb} is given by the following equations:
\begin{align}
    &\frac{1}{\mathcal{N}} \left[\sum_{\bk}\Pi_if\left(\R t_{\bk}\R^{\dagger}+\lambda\right)\Pi_i\right]_{ba} = \left[\Delta_i\right]_{ab}
    \label{detDelta}
    \\
    &\frac{1}{\mathcal{N}}\left[\sum_{\bk}\Pi_i 
    t_{\bk}\R^{\dagger} f\left(\R t_{\bk}\R^{\dagger}+\lambda\right)\Pi_i\right]_{\alpha a} = \sum_{c,a=1}^{\tilde{\nu}_i}\sum_{\alpha=1}^{\nu_i}\left[\D_i\right]_{c\alpha}\left[\Delta_i\left(1-\Delta_i\right)\right]^{\tfrac{1}{2}}
    \label{detD}
    \\
   &\sum_{c,b=1}^{\tilde{\nu}_i}\sum_{\alpha=1}^{\nu_i}\frac{\partial}{\partial \left[d^0_i\right]_s} \left(\left[\Delta_i\left(1-\Delta_i\right)\right]^{\tfrac{1}{2}}_{cb}\left[\D_i\right]_{b\alpha}\left[\R_i\right]_{c\alpha} + \mathrm{c.c.}\right) + [l_i+l^c_i]_{s} = 0
   \label{detLc}
   \\
   &\h_i^{\mathrm{emb}}\ket{\Phi_i} = E_i^c\ket{\Phi_i}
   \\
   &\left[\F_i^{(1)}\right]_{\alpha a} = \bra{\Phi_i}\hat{c}^{\dagger}_{i\alpha}\hat{f}^{\phantom{\dagger}}_{ia}\ket{\Phi_i}
   - \sum_{c=1}\left[\Delta_i\left(1-\Delta_i\right)\right]^{\tfrac{1}{2}}\left[\R_i\right]_{c\alpha} \overset{!}{=} 0
   \label{detF1}
   \\
   &\left[\F_i^{(2)}\right]_{ab} = \bra{\Phi_i}\hat{f}^{\phantom{\dagger}}_{ib}\hat{f}^{\dagger}_{ia}\ket{\Phi_i} - \left[\Delta_i\right]_{ab} \overset{!}{=} 0
   \,,
   \label{detF2}
\end{align}
where $f$ is the zero-temperature Fermi function and in Eqs.~\eqref{detDelta},\eqref{detD} we introduced the following block-matrices:
\begin{align}
   \label{blockL}
   \lambda = \begin{pmatrix}
    \left[\mathbf{0}\right]_{\nu_0\times\nu_0} & \dots & \dots & \mathbf{0} \\
    \vdots & \lambda_1 &  \vdots \\
    \vdots & \vdots &  \ddots & \vdots \\
    \mathbf{0} & \dots &  \dots & \lambda_M
  \end{pmatrix} 
  \,,
\end{align}

\begin{align}
   \label{blockR}
   \R = \begin{pmatrix}
    \left[\mathbf{1}\right]_{\nu_0\times\nu_0}  &\dots& \dots & \mathbf{0} \\
    \mathbf{0} & \R_1 & \dots & \vdots \\
    \vdots  & \vdots  & \ddots & \vdots \\
    \mathbf{0} & \dots& \dots & \R_M
  \end{pmatrix}
  \,,
\end{align}
where $\left[\mathbf{1}\right]_{n\times n}$ is the $n\times n$ identity matrix
and $\left[\mathbf{0}\right]_{n\times n}$ is the $n\times n$ zero matrix.
We also introduced the following projectors:
\begin{align}
   \label{Proj}
   \Pi_i = \begin{pmatrix}
    \delta_{i0}\left[\mathbf{1}\right]_{\nu_0\times\nu_0} & \dots &\dots & \mathbf{0} \\
    \vdots & \delta_{i1}\left[\mathbf{1}\right]_{\tilde{\nu}_1\times\tilde{\nu}_1} & \dots & \vdots \\
    \vdots & \vdots  & \ddots & \vdots \\
    \mathbf{0}  & \dots& \dots & \delta_{iM}\left[\mathbf{1}\right]_{\tilde{\nu}_M\times\tilde{\nu}_M}
  \end{pmatrix}
  \,.
\end{align}
Finally, we introduced the following expansions of the matrices $\Delta_i$, $\lambda_i$, $\lambda^c_i$, in terms of an orthonormal
basis of Hermitian matrices $\left\{\left[h_i\right]_s\right\}$ (with respect to the canonical scalar product $(A, B) = \Tr \left[A^{\dagger}B\right]$):
\begin{align}
    \label{coeffDelta}
    \Delta_i =& \sum_{s=1}^{\tilde{\nu}_i^2} \left[d^0_i\right]_s {}^t\left[h_i\right]_s \\
    \label{coeffL}
    \lambda_i =& \sum_{s=1}^{\tilde{\nu}_i^2} \left[l_i\right]_s \left[h_i\right]_s  \\
    \label{coeffLc}
    \lambda^c_i =& \sum_{s=1}^{\tilde{\nu}_i^2} \left[l^c_i\right]_s \left[h_i\right]_s \,,
\end{align}
where $\left[d^0_i\right]_s$, $\left[l_i\right]_s$ and $\left[l^c_i\right]_s$ are real-valued coefficients.

\subsection{Algorithmic structure of the g-GA}
The equations \eqref{detDelta}-\eqref{detF2} can be solved numerically using the following numerical procedure (represented schematically in Fig.~\ref{FigureS2}):
\begin{enumerate}
    \item Starting from an initial guess for the coefficients $\left[r_i\right]_a$ and $\left[l_i\right]_s$ construct the matrix $\R_i$ directly by using $\left[r_i\right]_a$ as the matrix elements of $\R_i$ and the matrix $\lambda_i$ according to Eq. \eqref{coeffL}.
    \item Arrange the $\R_i$ and $\lambda_i$ in block matrices over all $i$ as shown in Eqs. \eqref{blockL} and \eqref{blockR} and construct $\h_{\mathrm{qp}}$ as shown in Eq.\eqref{hqp}.
    \item Compute $\Delta_i$ according to Eq. \eqref{detDelta} using the projector $\Pi_i$ as given in Eq. \eqref{Proj}.
    \item Compute the Lagrange multipliers $\D_i$ by inverting Eq. \eqref{detD}.
    \item Determine the coefficients $\left[l^c_i\right]_s$ from Eq. \eqref{detLc} and construct the matrix $\lambda^c_i$ according to Eq. \eqref{coeffLc}.
    \item Construct $\h^{\mathrm{emb}}_i$ according to Eq.~\eqref{hemb} and calculate its ground state $\ket{\Phi_i}$ within the subspace with $(\nu_i+\tilde{\nu}_i)/2$ electrons (i.e., half of the total number of EH modes).
    \item Compute $\F_i^{(1)}$ and $\F_i^{(2)}$ according to equations \eqref{detF1} and \eqref{detF2}.
\end{enumerate}
A saddle-point of $\Lag$ satisfies $\F_i^{(1)}=\F_i^{(2)}=0$, i.e.,
the procedure described above has to be repeated until a self-consistent solution to these equations is found.

\subsection{Gauge invariance of the g-GA} \label{sec:Gauge}

Following Ref. \onlinecite{Lanata2016} it can be readily shown that also the g-GA Lagrangian is invariant with respect to the following Gauge transformation:
\begin{align}
    \ket{\Psi_0} &\rightarrow \U^{\dagger}\left(\theta\right)\ket{\Psi_0} 
    \label{g1}
    \\
    \ket{\Phi_i} &\rightarrow U_i^{\dagger}\left(\theta_i\right)\ket{\Phi_i} \\
    \R_i &\rightarrow u^{\dagger}_i\left(\theta_i\right)\R_i \\
    \D_i &\rightarrow {}^t u\left(\theta_i\right)\D_i \\
    \Delta_i &\rightarrow {}^t u_i\left(\theta_i\right)\Delta_i {}^t u_i^{\dagger}\left(\theta_i\right) \\
    \lambda_i &\rightarrow u_i^{\dagger}\left(\theta_i\right)\lambda_i u_i \left(\theta_i\right) \\
    \lambda_i^c &\rightarrow u_i^{\dagger}\left(\theta_i\right) \lambda_i^c u_i\left(\theta_i\right)
    \label{g7}
    \,,
\end{align}
with:
\begin{align}
    u_i\left(\theta_i\right) &= e^{i\theta_i} \\
    U_i\left(\theta_i\right) &= e^{i\sum_{a,b=1}^{\tilde{\nu}_i}\left[\theta_i\right]_{ab} \hat{f}^{\dagger}_{ia}\hat{f}^{\phantom{\dagger}}_{ib}} \\
    \U\left(\theta\right) &= e^{i\sum_{R}\sum_{i\geq 1}\sum_{a,b=1}^{\tilde{\nu}_i}\left[\theta_i\right]_{ab} \fc_{Ria}\fa_{Rib}} \,,
\end{align}
where $u_i\left(\theta_i\right)\in\mathbb{C}^{\tilde{\nu}_i\times\tilde{\nu}_i}$, $U_i\left(\theta_i\right)\in\mathbb{C}^{2^{\tilde{\nu}_i}\times 2^{\tilde{\nu}_i}}$
and $\U\left(\theta\right)\in\mathbb{C}^{2^{\tilde{\nu}}\times 2^{\tilde{\nu}}}$ and $\theta_i$ are Hermitian matrices.

\subsection{General analytical expression for spectral function from g-GA variational parameters}

Let us consider the g-GA zero-temperature spectral function, defined as follows:
\be
\mathcal{A}_{i\alpha,j\beta}(\bk,\omega)=
\Av{\Psi_G}{\ca_{\bk i \alpha}\,\delta(\omega-\h)\,\cc_{\bk j \beta}}+
\Av{\Psi_G}{\cc_{\bk j \beta}\,\delta(\omega+\h)\,\ca_{\bk i \alpha}}\,.
\label{Ghost-Ak}
\ee
Following Ref.~\onlinecite{Ghost-GA}, we obtain the
following approximation to the physical Green's function:
\be
\mathcal{G}_{i\alpha,j\beta}(\bk,\omega)=\int_{-\infty}^{\infty}d\epsilon\,
\frac{\mathcal{A}_{i\alpha,j\beta}(\bk,\omega)}{\omega-\epsilon}
\simeq\left[
\R^\dagger_i\,\Pi_i\frac{1}{\omega-[\R \epsilon_{\bk}
	\R^\dagger+\lambda]}\Pi_j\,\R_j
\right]_{\alpha\beta}
\,.
\label{tG}
\ee

As explained above, 
the g-GA Lagrange function is invariant under the group of gauge transformations defined by Eqs.~\eqref{g1}-\eqref{g7}.
To calculate the self-energy, it is convenient to
choose a gauge such that $\lambda_i$ is diagonal and decompose $R_i$ and $\lambda_i$ as follows:
\begin{align}
    \R_i &= \begin{pmatrix} \R_i^{(1)} \\ \R_i^{(2)} \end{pmatrix} 
    \label{decR}
    \\
    \lambda_i &= \begin{pmatrix} \lambda_i^{(1)} & \mathbf{0} \\ \mathbf{0} & \lambda_i^{(2)} \end{pmatrix}
    \,,
    \label{lambdadec}
\end{align}
where $\R_i^{(1)}$ and $\lambda_i^{(1)}$ are matrices of dimension $\nu_i\times\nu_i$, while $\R_i^{(2)}$ and $\lambda_i^{(2)}$ are $(\tilde{\nu}_i-\nu_i)\times\nu_i$ and 
$(\tilde{\nu}_i-\nu_i)\times(\tilde{\nu}_i-\nu_i)$, respectively.
In this gauge, using Eqs.~\eqref{decR}, \eqref{lambdadec} and the Dyson equation:
\begin{align}
    \G(\bk,\omega) = \left[\omega - \tau_{\bk}-\Sigma(\omega)\right]^{-1}
\end{align}
(where $\tau_{\bk} = t_{\bk} + t^{\mathrm{loc}}$
also contains all local terms of Eq.~\eqref{hloc-quadratic}),
it can be shown with a straightforward calculation that the self-energy is given by the following equation:
\begin{align}
   \Sigma(\omega) = \begin{pmatrix}
    \left[\mathbf{0}\right]_{\nu_0\times\nu_0} &\mathbf{0}& \dots & \mathbf{0} \\
    \mathbf{0} & \Sigma_1(\omega) & \dots &\vdots \\
    \vdots & \vdots  & \ddots & \vdots \\
    \mathbf{0} & \dots& \dots & \Sigma_M(\omega)
  \end{pmatrix}
  \,,
\end{align}
where, as expected, $\Sigma(\omega)$ is local ($\bk$-independent) and
$\Sigma_i(\omega)$ are the $\nu_i\times\nu_i$ matrices $\forall\,i\geq 1$, given by the following equation:
\begin{align}
    \Sigma_{i}(\omega) =& t^{\mathrm{loc}}_i -\omega\frac{\mathbf{1}-\R_i^{(1)\dagger}\R_i^{(1)}}{\R_i^{(1)\dagger}\R_i^{(1)}} +\left[\R_i^{(1)}\right]^{-1} \lambda^{(1)}_i\left[\R_i^{(1)\dagger}\right]^{-1} + \left[\R_i^{(1)}\right]^{-1}\left(\omega-\lambda_i^{(1)}\right)\left[\R_i^{(1)\dagger}\right]^{-1}\R_i^{(2)\dagger} \nonumber\\ 
    &\times
    \Biggl[\R_i^{(2)}\left[\R_i^{(1)}\right]^{-1}\left(\omega-\lambda^{(1)}_i\right)\left[\R_i^{(1)\dagger}\right]^{-1}\R_i^{(2)\dagger}+\left(\omega-\lambda_i^{(2)}\right)\Biggr]^{-1}
    \nonumber\\
    &\times\R_i^{(2)}\left[\R_i^{(1)}\right]^{-1} \left(\omega-\lambda_i^{(1)}\right)\left[\R_i^{(1)\dagger}\right]^{-1}
    \,,
\end{align}
which reduces to Eq.~10 of the main text for single-orbital impurities.

Note that, as pointed out in the main text, from Eq.~\eqref{tG} it follows that the poles of the Green's function lie on top of the eigenvalues of the quasi-particle Hamiltonian, see Eq.~\eqref{hqp}.
However, because of the matrices $\mathcal{R}^\dagger_i$ and $\mathcal{R}_j$ at the left and the right of Eq.~\eqref{tG} ---which are rectangular, ---
only a portion of the quasi-particle spectral weight is physical (see Fig.~3 of the main text).

\section{Pathology of the bare GA in the narrow-bandwidth limit of the ALM}

\begin{figure} 
    \includegraphics[width=11.7cm]{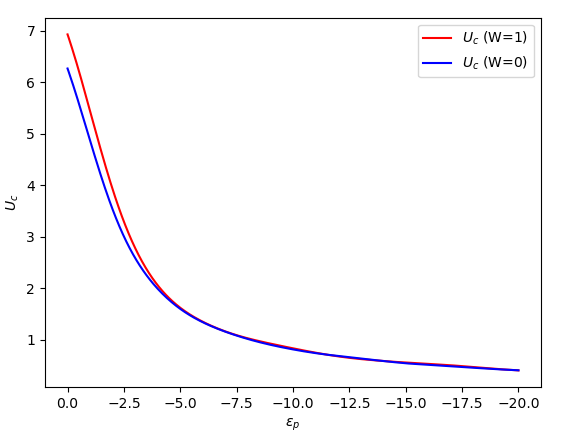}
    \caption{Behavior of $U_c$ as a function of $\epsilon_p$ in bare GA for $V=1$, both for $W=1$ (red) and in the narrow-bandwidth limit $W=0$ (blue).
    }
    \label{FigureS3}
\end{figure}

As pointed out in the main text, within the GA the Mott transition of the ALM occurs (by construction) only when
$\langle\dc_{i\sigma}\pa_{i\sigma}
    \rangle_{\text{GA}}
    =r 
    \Av{\Psi_0}{\fc_{i\sigma}\pa_{i\sigma}}=0$,
i.e., when the variational parameter $r$ vanishes.
However, as shown in Fig.~1 of the main text, the GA critical value $U_c$ is overestimated dramatically for the ALM, especially for $\epsilon_p\ll -1$.

To explain this behavior, it is insightful to inspect the GA solution also in the narrow-bandwidth limit, i.e., setting the half-bandwidth $W$ of the Bethe-lattice hopping matrix to $0$ ---corresponding to a series of decoupled $p$-$d$ dimers.
The resulting evolution of $U_c$ (i.e., the interaction strength such that $r$ vanishes) is shown here in Fig.~\ref{FigureS3}, where we set
$V=1$ and consider $\epsilon_p\ll -1$.

We observe that, not surprisingly (since $V=1$),
$\langle\dc_{i\sigma}\pa_{i\sigma} \rangle_{\text{GA}}$ does not vanish at $U\sim 0$ (neither for $W=0$ nor for $W=1$).
In fact, at $U=0$ the $GA$ wavefunction is exact.
On the other hand, in this limit any infinitesimal $U$ should induce a Mott transition for $W\rightarrow 0$.
This simple observation showcases very clearly how, as pointed out in the main text, the $p$-$d$ charge fluctuations cannot be neglected in the Mott phase ---as they \emph{must} coexist in the narrow-bandwidth limit of the ALM.
In other words, the argument above shows that ---for small $W$--- the GA behavior of $U_c$ is essentially unrelated with the Mott transition.
We also note that, for the values of $\epsilon_p$ considered in Fig.~\ref{FigureS3}, the GA behavior of $U_c$ is almost independent of whether $W=0$ or $W=1$.
In fact, one can readily demonstrate also analytically that the bare GA predicts that $U_c(W)\sim U_c(W=0)\sim V^{2}/\left|\epsilon_{p}\right|$ $\forall\,W$ for $\epsilon_p\rightarrow -\infty$ (which is clearly incorrect).
%
This shows that the pathology of the GA, that we discussed above in the narrow-bandwidth limit, cannot be only regarded as an accident specific to this particular  case,
as it affects the GA predictions also for finite $W$.

In summary, while the bare GA captures some qualitative features of the phase diagram of the ALM, here we identified a pathological behavior of this method in the narrow-bandwidth limit, which leads to a dramatic overestimation of $U_c$ (especially for $\epsilon_p\rightarrow -\infty$).
From a general perspective, this problem can be traced back to the inability of the GA of describing simultaneously the Mott physics and the $p$-$d$ charge fluctuations.
In the main text we showed that the g-GA does not have such limitation and, consequently, it resolves the problems of the bare GA here uncovered, providing us with a description of the ALM with accuracy comparable to DMFT.




%